\documentclass[12pt]{iopart}
\usepackage{iopams}
\usepackage{graphicx}
\usepackage{upgreek}      
\usepackage[usenames,dvipsnames]{xcolor}
\usepackage{hyperref}
\usepackage{siunitx}
\usepackage{verbatim}
\usepackage{xspace}
\usepackage{lineno}

\hypersetup{
    colorlinks=true,
    linkcolor=cyan,
    filecolor=magenta,      
    urlcolor=blue,
    citecolor=blue,
}

\newenvironment{bmatrix}{\left[\begin{array}{cccccccc}}{\end{array}\right]}
\newcommand{\GeV}{\ensuremath{\,\text{Ge\hspace{-.08em}V}}\xspace}

\newcommand{\pT}{\ensuremath{p_{\mathrm{T}}}\xspace}
\newcommand{\QKeras}{{\textsc{QKeras}}\xspace}
\newcommand{\Keras}{{\textsc{Keras}}\xspace}
\newcommand{\TensorFlow}{{\textsc{TensorFlow}}\xspace}
\newcommand{\GEANTfour} {{\textsc{Geant4}}\xspace}
\newcommand{\RMSeff}{\ensuremath{\text{RMS}_\text{eff}}\xspace}

\begin{document}

\begin{flushright}
\footnotesize
FERMILAB-PUB-23-288-CMS-CSAID
\end{flushright}

\title[Differentiable EMD for Data Compression at the HL-LHC]{
    Differentiable Earth Mover's Distance for Data Compression at the High-Luminosity LHC
}
\author{
    Rohan Shenoy$^1$,
    Javier Duarte$^1$,
    Christian Herwig$^2$,
    James Hirschauer$^2$,
    Daniel Noonan$^2$,
    Maurizio Pierini$^3$,
    Nhan Tran$^2$, and
    Cristina Mantilla Suarez$^2$
}
\address{$^1$Department of Physics, University of California at San Diego, La Jolla, CA 92093,\\
    $^2$Particle Physics Division, Fermi National Accelerator Laboratory, Batavia, IL 60510,\\
    $^3$Experimental Physics Department, European Organization for Nuclear Research (CERN), 1211 Geneva 23, Switzerland}

\ead{rsshenoy@ucsd.edu,jduarte@ucsd.edu}
\vspace{10pt}
\begin{indented}
    \item[]22 December 2023
\end{indented}

\begin{abstract}
    The Earth mover's distance (EMD) is a useful metric for image recognition and classification, but its usual implementations are not differentiable or too slow to be used as a loss function for training other algorithms via gradient descent.
    In this paper, we train a convolutional neural network (CNN) to learn a differentiable, fast approximation of the EMD and demonstrate that it can be used as a substitute for computing-intensive EMD implementations.
    We apply this differentiable approximation in the training of an autoencoder-inspired neural network (encoder NN) for data compression at the high-luminosity LHC at CERN.
    The goal of this encoder NN is to compress the data while preserving the information related to the distribution of energy deposits in particle detectors.
    We demonstrate that the performance of our encoder NN trained using the differentiable EMD CNN surpasses that of training with loss functions based on mean squared error.
\end{abstract}

\submitto{\MLST}
\maketitle

\section{Introduction}
\label{sec:intro}
The Earth mover's distance (EMD)~\cite{emdimage}, also known as the 1st Wasserstein distance, has found use in various domains of machine learning (ML) problems.
The EMD is a function of optimal transport between two distributions of points $p_i$ with weights $w_{p_i}$ and $q_j$ with weights $w_{q_j}$ defined as
\begin{eqnarray}
    \text{EMD} = \frac{\sum_{i=1}^m \sum_{j=1}^n f_{ij}d_{ij}}{\sum_{i=1}^m \sum_{j=1}^n f_{ij}}
\end{eqnarray}
where $f_{ij}$ is the optimal flow between $p_i$ and $q_j$ that minimizes the overall cost of transportation and $d_{ij}$ is the ground distance between $p_i$ and $q_j$.
Roughly speaking, it measures the minimum work required to transport one distribution to another
It has been used in image classification~\cite{emd_classification}, visual tracking~\cite{tracking}, and medicine \cite{medical}.
In Ref.~\cite{Komiske2019}, the authors develop a generalization to the metric known as the energy mover's distance.
The energy mover's distance has been used to explore the geometry of particle jets~\cite{jetscmsod}, as a basis for defining event isotropy~\cite{isotropy}, to train variational autoencoders~\cite{emd_vae}, for anomaly detection~\cite{anomaly}, and to evaluate particle physics simulations~\cite{raghav}.

Two issues commonly arise from implementations of the EMD metric: (1) they are not differentiable and (2) they are often computationally intensive.
Efficient, differentiable, or linearized approximations to the EMD, have been studied in the literature to make them more suitable for a variety of tasks.
Ref.~\cite{deepemd} adopts the EMD as a metric to determine image relevance.
The authors use a structured fully connected layer with the energy mover's distance inserted as a layer (with its Jacobian computed using the implicit function theorem) to classify dense image representations.
However, the implementation in Ref.~\cite{deepemd} is slow and does not generalize to large datasets.
Another study~\cite{Cai:2020vzx} develops a linearized approximation to the energy mover's distance.
Ref.~\cite{steven} proposes the use of graph neural networks (GNNs) to approximate the energy mover's distance for use in particle reconstruction.
The Sinkhorn distance~\cite{sinkhorn} has also been proposed as faster, differentiable (upper bound) approximation to EMD, and has been used in particle physics applications~\cite{Ba:2023hix}.
However, in some applications, including the one we study, this approximation can be inaccurate.
Recently, the authors of Ref.~\cite{Kitouni:2022qyr} developed a way to approximate the energy mover's distance in a differentiable way using a Lipschitz network~\cite{Kitouni:2021fkh} that enforces an exact upper bound on the Lipschitz constant of the model by constraining its weights.
Ref.~\cite{Larkoski:2023qnv} introduced a new one-dimensional encoding of collider events, which allows for a closed-form expression of the 1st Wasserstein distance dubbed the spectral energy mover's distance.
Ref.~\cite{gouskos2023optimal} uses a ``sliced" Wasserstein distance~\cite{wass1} as a loss function for training an attention-based GNN for pile-up mitigation.

In this paper, we propose the use of convolutional neural networks (CNNs) to learn a differentiable approximation of the EMD metric.
We then use this metric to optimize an algorithm for front-end data compression at the high-luminosity LHC (HL-LHC).
In particular, we train a convolutional autoencoder to directly minimize our differentiable approximation to the EMD via gradient-based optimizers.
We demonstrate that this technique leads to an autoencoder that better preserves the relevant physical information stored in the shape of the energy deposits created by electrons and other high energy particles.
The rest of this paper is organized as follows.
Section~\ref{sec:hgcal} introduces the high-granularity calorimeter (HGCAL), and the energy concentrator trigger (ECON-T), which are the setting for data compression example.
In Section~\ref{sec:EMDCNN}, we introduce the EMD CNN and its optimization.
Section~\ref{sec:aeEMD} demonstrates the use of the EMD as a loss function for the ECON-T autoencoder, and in Section~\ref{sec:physics} we show an improved trigger performance.
We summarize our results in Section~\ref{sec:summary}.

\section{HGCAL Autoencoder}
\label{sec:hgcal}
The next generation of big data science experiments, including the HL-LHC at CERN, presents challenges because of the massive data rates generated during data taking.
One solution is to immediately compress data before transmitting it off the detector in front-end electronics.
A deep learning technique, based on training an autoencoder, has recently been proposed~\cite{ASIC} to compress data from a new HGCAL~\cite{HGCALTDR} that is part of the upgraded CMS detector.
The HGCAL includes over 6 million readout channels that must be compressed before being sent to the level-1 trigger (L1T), a real-time filter system implemented on field-programmable gate arrays (FPGAs)~\cite{CMSP2L1T}.
In high energy physics experiments, the L1T~\cite{CMSL1T,CMSP2L1T,ATLASL1T,ATLASP2L1T} is the first stage of the system that decides which collision events are interesting enough to record, such as events potentially containing a Higgs boson.
In order to provide input to the L1T, the HGCAL transmits a stream of trigger data at a frequency of 40\,MHz.
Application-specific integrated circuits (ASICs) are used to digitize and encode trigger data before transmission to back-end FPGAs for further processing.
The goal of the algorithm is to compress the HGCAL data on an ASIC while preserving the information related to the distribution of energy deposits in silicon particle detector sensors.

\begin{figure}[ht]
\label{fig:asic}
    \includegraphics[width=0.96\textwidth]{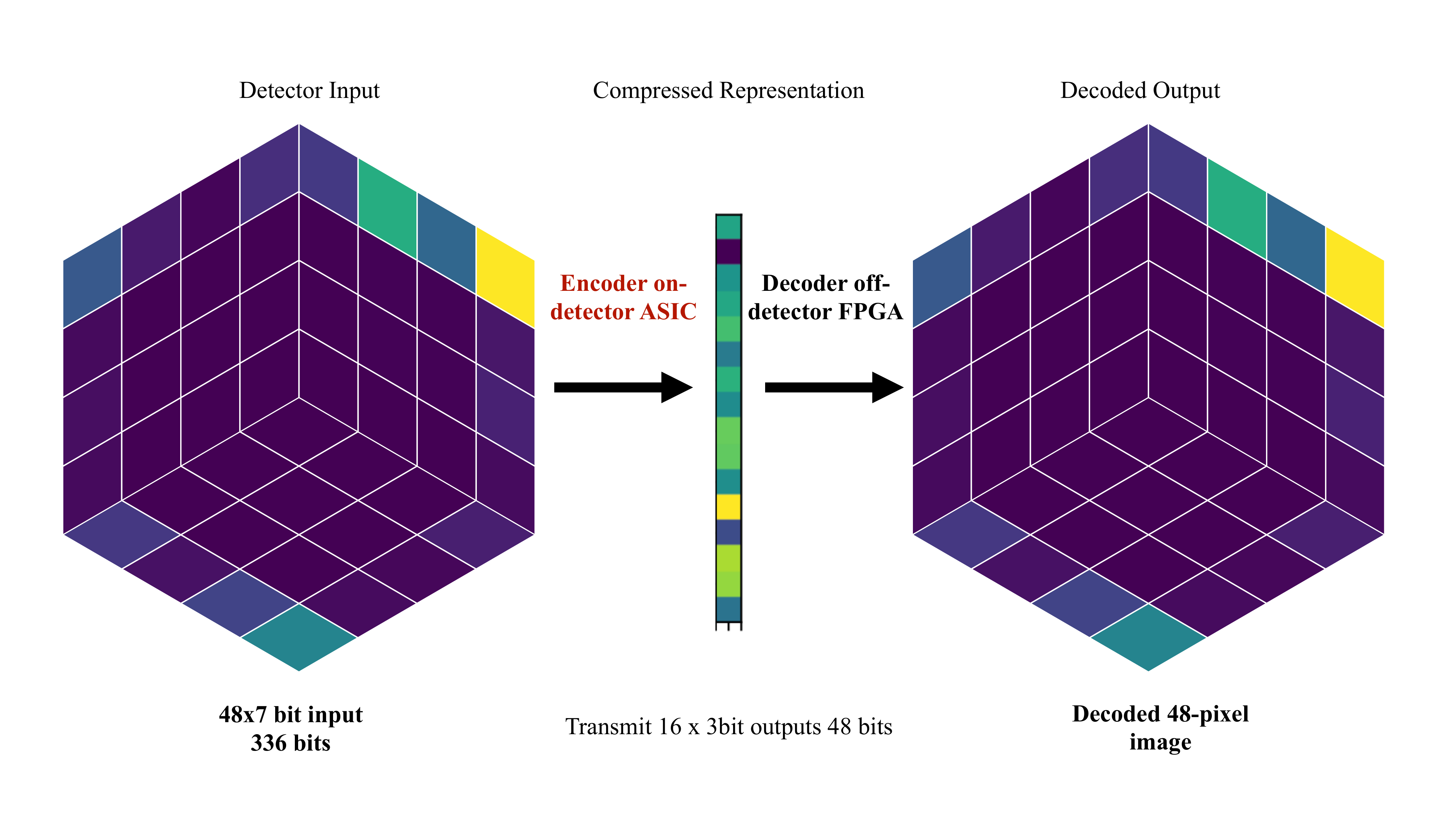}
    \caption{Conceptual overview of the HGCAL trigger path for the autoencoder.
        We read a 7 bit input for each of the 48 trigger cells, and compress it to a latent representation of 16$\times$3 bits.
        The decoder mirrors the encoder to produce the original 48 input image.}
\end{figure}

The ASIC autoencoder (composed of an encoder and decoder neural network) is trained to reproduce the input data while producing an optimum latent representation.
The encoder is a reconfigurable CNN trained with \QKeras, a quantization-aware training library~\cite{Coelho:2020zfu}.
While the architecture of this encoder will remain fixed, the weights will be reprogrammable.
Ref.~\cite{ASIC} details the hardware implementation of the encoder onto the ASIC.
Groups of 2$\times$2 or 3$\times$3 HGCAL readout channels are summed together into trigger cells (TCs).
Each HGCAL silicon sensor consists of 48 such TCs, with each TC transmitting 7 bits of data at 40\,MHz.
These 48 TCs are arranged in the hexagonal geometry as shown in Figure~\ref{fig:asic}.
The task of the ECON-T ASIC is to aggregate the 48 7-bit signals and compress them into a representation that allows for the best decision making by the L1 trigger.
While the range of bits in the output varies per expected cell occupancy, this study is conducted at the maximum expected cell occupancy: 9 bits for each encoded value.

\subsection{Baseline Loss Functions}
\label{sec:losses}
As a proxy for the ultimate physics performance, the EMD between the HGCAL input and the decoded autoencoder output can be used as a metric to evaluate a given model.
As mentioned in Section~\ref{sec:intro}, this EMD metric as implemented in the Python Optimal Transport~\cite{flamary2021pot} library cannot directly be optimized using for gradient-based methods.
Therefore, alternative loss functions have been used in the past for the optimization of the autoencoder.
We compare the performance of the autoencoder when trained with three different loss functions: the EMD CNN approximation we introduce in section~\ref{sec:EMDCNN}, and two additional baselines described here.

The first baseline for comparison is a simple weighted mean-squared error (MSE) loss function
\begin{equation}
    L_{\text{Weighted MSE}} = \frac{1}{N_{1\times1}}\sum\limits_{i=0}^{N_{1\times1}}(x_i - y_i)^2\max(x_i, y_i),
    \label{eqn:weightedMSE0}
\end{equation}
where $N_{1\times1}=48$ is the number of cells, $x_i$ is the set of charge fractions input to the encoder, and $y_i$ is the set of charge fractions output by the decoder.
The square of the differences is weighted by the larger of the two charge fractions, with the aim of emphasizing the accuracy of trigger cells containing the most charge.

The second, improved baseline is an ad-hoc loss function proposed in Ref.~\cite{ASIC} known as the telescope MSE loss function.
This loss function is designed to better capture differences in the shape of the spatial distribution of charge deposits.
The telescope MSE loss function computes a weighted MSE for multiple groupings of trigger cells, at different ``telescoping'' granularities: individual cells (denoted 1$\times$1), 2$\times$2 sums of trigger cells, and 4$\times$4 sums of trigger cells.

\begin{figure}[ht]
    \centering
    \includegraphics[width = 0.8\linewidth]{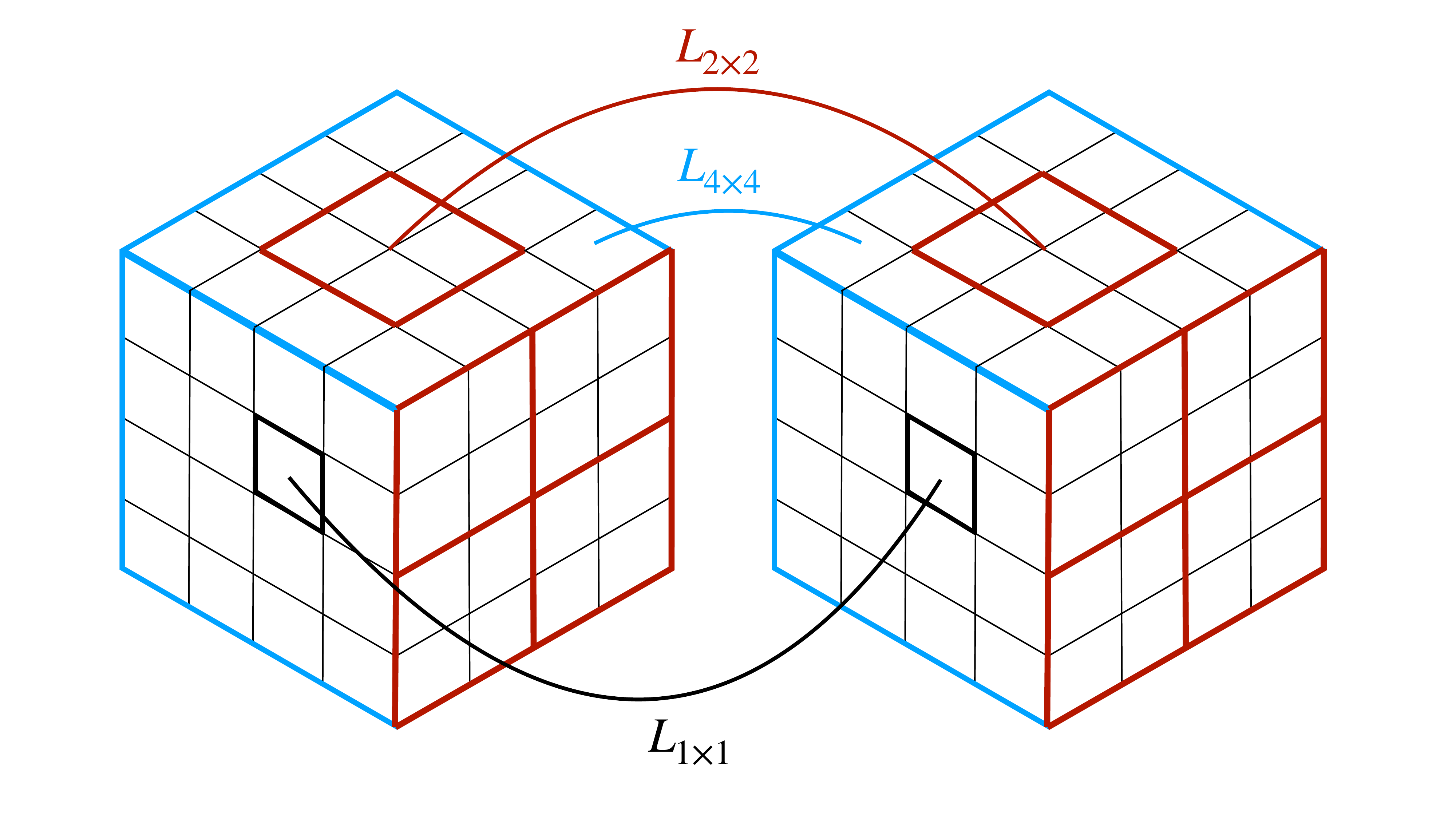}
    \caption{
        Illustration of the telescope loss.
        The red lines correspond to one example of a 2$\times$2 trigger cell grouping in the $L_{2\times2}$ loss term, while the blue lines correspond to the three 4$\times$4 trigger cell groupings in the $L_{4\times4}$ loss term for telescope loss.}
    \label{fig:tele}
\end{figure}

The component of the loss from the individual cells $L_{1\times1}$ is equal to $L_{\text{Weighted MSE}}$ from Eq.~(\ref{eqn:weightedMSE0}).
For calculating the contribution from 2$\times$2 sums of trigger cells, a weighted MSE is found from charge fractions of neighboring 2$\times$2 groupings of trigger cells.
To account for the fact that cells along the edge of the wafer will not be included in as many 2$\times$2 sums, an additional factor is applied to each weight based on the number of times its constituent trigger cells get counted into sums.
The component of the telescope MSE function from the 2$\times$2 sums, $L_{2\times2}$, is defined as
\begin{equation}
    L_{2\times2} = \frac{1}{N_{2\times2}}\sum\limits_{i=0}^{N_{2\times2}}w_i(x^{2\times2}_i - y^{2\times2}_i)^2\max(x^{2\times2}_i, y^{2\times2}_i),
    \label{eqn:telescope_2x2}
\end{equation}
where $x^{2\times2}_i = \sum_{j\in S^{2\times2}_i} x_j$ is the sum of the input trigger cells in the $i$th $2\times2$ grouping $S^{2\times2}_i$, $y^{2\times2}_i$ is the $i$th $2\times2$ sum of the output of the decoder, and $w_i$ is the weight corresponding to the $i$th sum.
To account for the fact that cells along the edge of the wafer will not be included in as many 2$\times$2 sums, an additional factor is applied to each weight based on the number of times its constituent trigger cells are included in sums.
For example, a 2$\times$2 sum can be formed from $q_0$, $q_1$, $q_8$ and $q_9$ (see Fig.~\ref{fig:tcArrange} for the numbering scheme).
Based on their locations on the edge of the detector, $q_0$ is only included in one sum, $q_1$ and $q_8$ are included in 2 sums, and $q_9$ is included in 4 separate sums.
The sum made up of these four trigger cells has a weight of $w_i = (1/1 + 1/2 + 1/2 + 1/4)/4 = 2.25/4$, with the extra factor of $1/4$ included to keep the scale the 2$\times$2 component of the loss function consistent with the individual cells because each charge fraction can be included in up to 4 sums.
Table~\ref{tab:2x2sums} in \ref{sec:table} shows the $N_{2\times2}=36$ 2$\times$2 sums and corresponding weights.

Finally, three 4$\times$4 sums are produced.
The trigger cells that get combined into the 4$\times$4 sums are shown in blue in Fig.~\ref{fig:tele}.
Because trigger cells are included in one and only one 4$\times$4 sum, no additional weight term is necessary here.
Thus, the component of the telescope MSE from the 4$\times$4 sums, $L_{4{\times}4}$, is given by
\begin{equation}
    L_{4{\times}4} =\frac{1}{N_{4{\times}4}} \sum\limits_{i=0}^{N_{4{\times}4}}(x^{4\times4}_i - y^{4\times4}_i)^2\max(x^{4{\times}4}_i, y^{4{\times}4}_i).
    \label{eqn:telescope_4x4}
\end{equation}
The total value of the telescope MSE loss is then a weighted sum of the three components
\begin{equation}
    \text{Telescope MSE} = 4 L_{1\times1} + 2L_{2\times2} + L_{4\times4}.
    \label{eqn:telescopeMSE}
\end{equation}
Weighting the individual loss terms in this fashion corresponds more closely to EMD.
The chosen weights are inversely related to the number of trigger cells in each grouping.

\subsection{Hexagonal Geometry Encoding}
\label{sec:encoding}
\begin{figure}[htbp]
    \centering
    \includegraphics[width=0.96\textwidth]{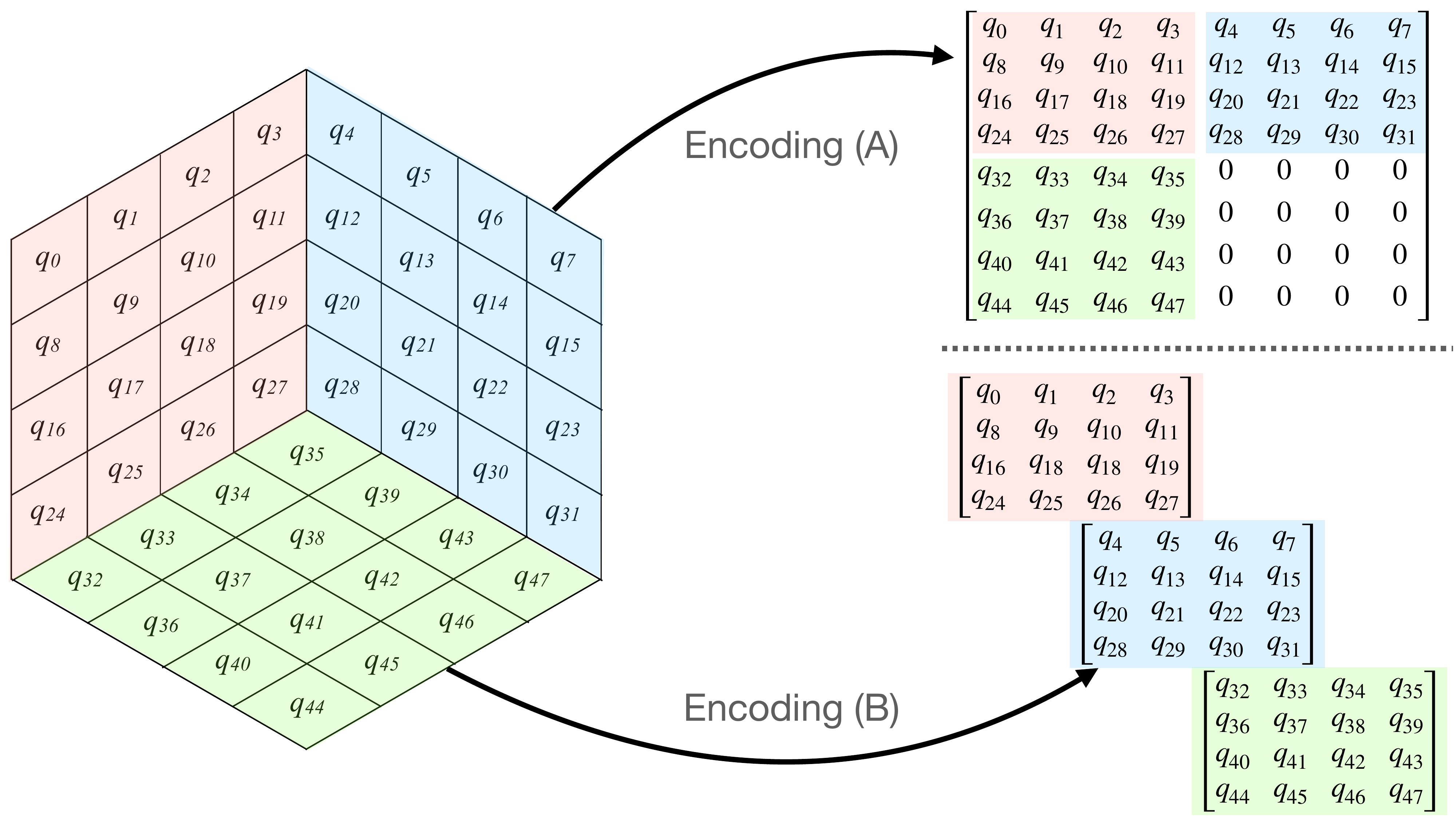}
    \caption{Remappings of the 48 trigger cell charge fractions into encoding (A): an 8$\times$8$\times$1 tensor with 16 empty cells, and encoding (B): a 4$\times$4$\times$3 tensor. \label{fig:tcArrange}}
\end{figure}

The HGCAL wafers have a unique hexagonal geometry because hexagonal wafers have an efficient spatial use of the silicon sensors~\cite{HGCALTDR}.
Conventional CNNs require regular/rectangular grids of pixels, possibly arranged in different channels, e.g., corresponding to red/blue/green in natural images.
There is no ideal remapping of these hexagonal wafers into rectangular grids that preserves all the nearest neighbors of all cells.
To account for this unique geometry, we use two remapping strategies.
Encoding (A) shown in Fig.~\ref{fig:tcArrange} (left) remaps the trigger cells into an $8{\times}8$ image with 1 channel, corresponding to the charge fraction in that cell.
This encoding is used for the autoencoder as it results in a smaller input size, which allows for a more efficient hardware implementation.
Encoding (B) shown in Fig.~\ref{fig:tcArrange} (right) remaps the trigger cells into a $4{\times}4$ image with three separate channels.
The three channels correspond to the charge fractions in three different areas of the hexagonal wafer.
It was chosen for the EMD CNN training to maintain the best performance as the same hardware constraints do not apply.
Once we develop a neural network loss with encoding (B), we remap to encoding (A) when we train the autoencoder.

\section{Approximating EMD with a Convolutional Neural Network}
\label{sec:EMDCNN}
\subsection{Dataset}

To train our neural networks we use simulated electron-positron events in HL-LHC conditions, corresponding to up to 200 additional proton-proton interactions within the same or nearby bunch crossings (pileup).
Pileup can contribute additional energy depositions in the HGCAL.
As the main goal of ECON-T is to identify and trigger on low-momentum electrons or positrons, we select events containing an electron (positron) with a maximum transverse momentum (\pT) of 35\GeV.
The dataset in ROOT format before filtering and preprocessing is published on the Zenodo platform~\cite{rohan_shenoy_2023_8338608,rohan_shenoy_2023_8408943}.

The detector response is modeled using a detailed description of the upgraded CMS detector based on \GEANTfour~\cite{GEANT4}.
The detector is described using a Cartesian coordinate system with the $z$ axis oriented along the beam axis and $x$ and $y$ axes defining the transverse plane.
The azimuthal angle $\phi$ is computed from the $x$ axis.
The polar angle $\theta$ is used to compute the pseudorapidity $\eta = -\log(\tan(\theta/2))$.

The target EMD for training is computed using the \texttt{ot.emd2} function from the Python Optimal Transport library~\cite{flamary2021pot}.
Generating a dataset for training the EMD CNN requires pairs of samples.
The total number of sample pairs in the dataset is 340\,000.
The dataset is split into 70\% for training and 30\% for validation.
We ensure that there is no overlap in the samples between the different datasets.

\subsection{Architecture and Hyperparameter Optimization}

We implement the neural network using the \Keras~\cite{chollet2015keras} API of \TensorFlow~\cite{tensorflow2015-whitepaper} with multiple convolutional and dense layers on the remapped 4$\times$4$\times$3 tensor input.
We average the output over both input orders $f(x_1, x_2) = \frac{1}{2}(g(x_1, x_2) + g(x_2, x_1))$ to enforce the symmetry in the EMD metric.
The model's hyperparameters are optimized with a grid search.
The evaluation metric we use to compare the performance of different models is the standard deviation of the distribution of the relative difference between the predicted and true EMDs $\sigma_\text{EMD rel. diff.}$ evaluated on the validation dataset. 
All trainings were performed on a NVIDIA GeForce GTX-1080 Ti GPU.

In the grid search, we varied multiple hyperparameters related to the number and sizes of the layers.
The number of 2D convolutional layers ranged from 1 to 4.
For these layers we tested kernel sizes of 1, 3, and 5, and 32, 64, 128, and 256 filters.
Each 2D convolutional layer is followed by a batch normalization layer~\cite{batchnorm} and rectified linear unit (ReLU) activation function~\cite{relu1,relu2}.
We tested three different loss functions: standard MSE, mean-squared logarithmic error (MSLE), and Huber~\cite{huber}.
However, we found that MSE loss function performed the best.
For each hyperparameter choice, the training was performed using the Adam optimizer~\cite{adam} with a learning rate of $1\times 10^{-3}$ for 120 epochs.
We found that one fully-connected layer was sufficient, but increasing the number of 2D convolutional layers improved our performance.
The best hyperparameters and corresponding $\sigma_\text{EMD rel. diff.}$ are detailed in \ref{sec:loss}.

\begin{figure}[ht]
    \centering
    \includegraphics[width=0.96\textwidth]{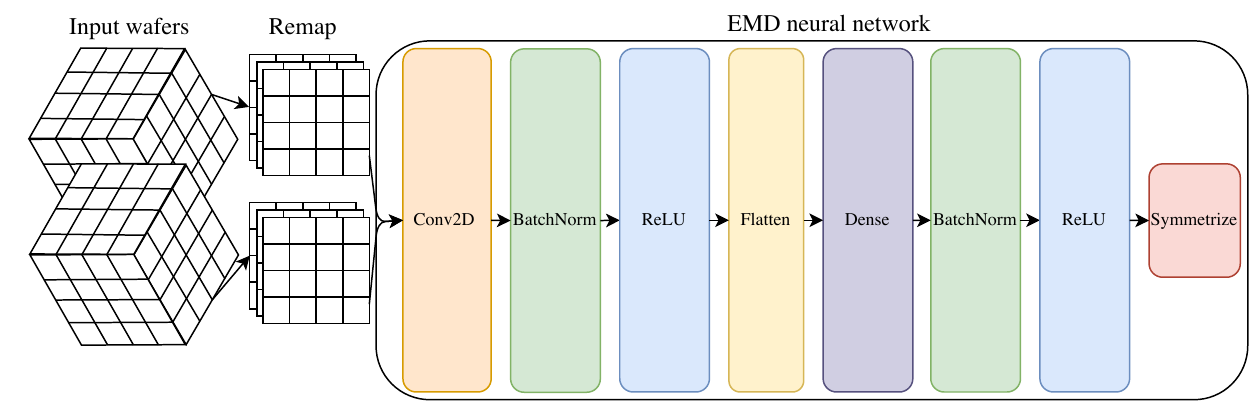}
    \label{fig:emd_nn}
    \caption{Architecture of our EMD CNN.
        We remap hexagonal HGCAL wafers to a more regular 4$\times$4$\times$3 tensor, and take in pairs of these remapped wafers as our input.
        The neural network has multiple 2D convolutional layers, each followed by a batch normalization layer and ReLU activation.
        The network then feeds into a single fully-connected layer, followed by a batch normalization layer and ReLU activation.
        We then average over both input orders to enforce the symmetry in the EMD metric.}
\end{figure}

The optimized model, shown in Fig.~\ref{fig:emd_nn} has four 2D convolutional layers, each with 32 filters, a kernel size of 5, and a stride of 1. There is one fully-connected layer with 256 neurons.

\begin{figure}[ht]
    \centering
    \includegraphics[width=0.45\columnwidth]{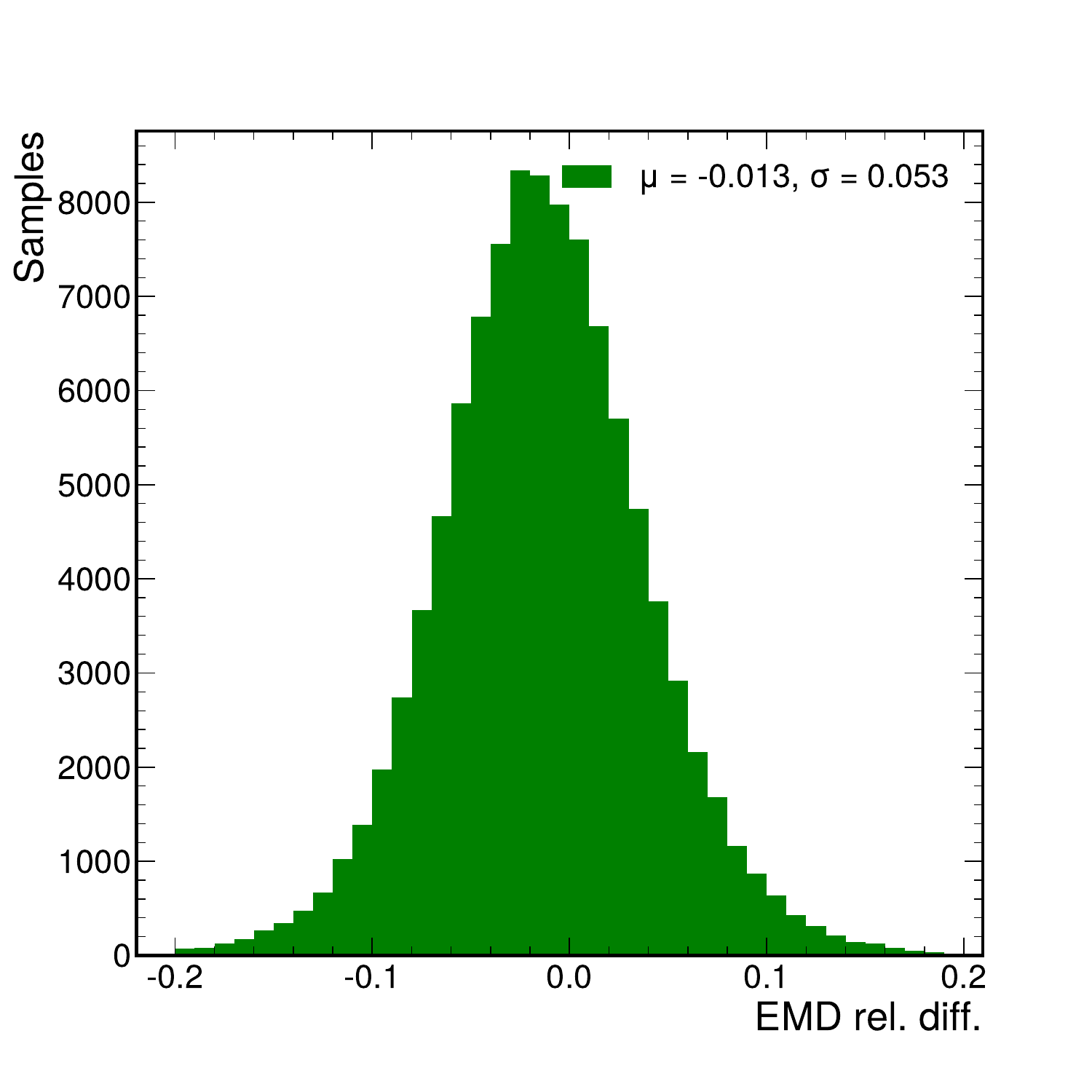}
    \includegraphics[width=0.45\columnwidth]{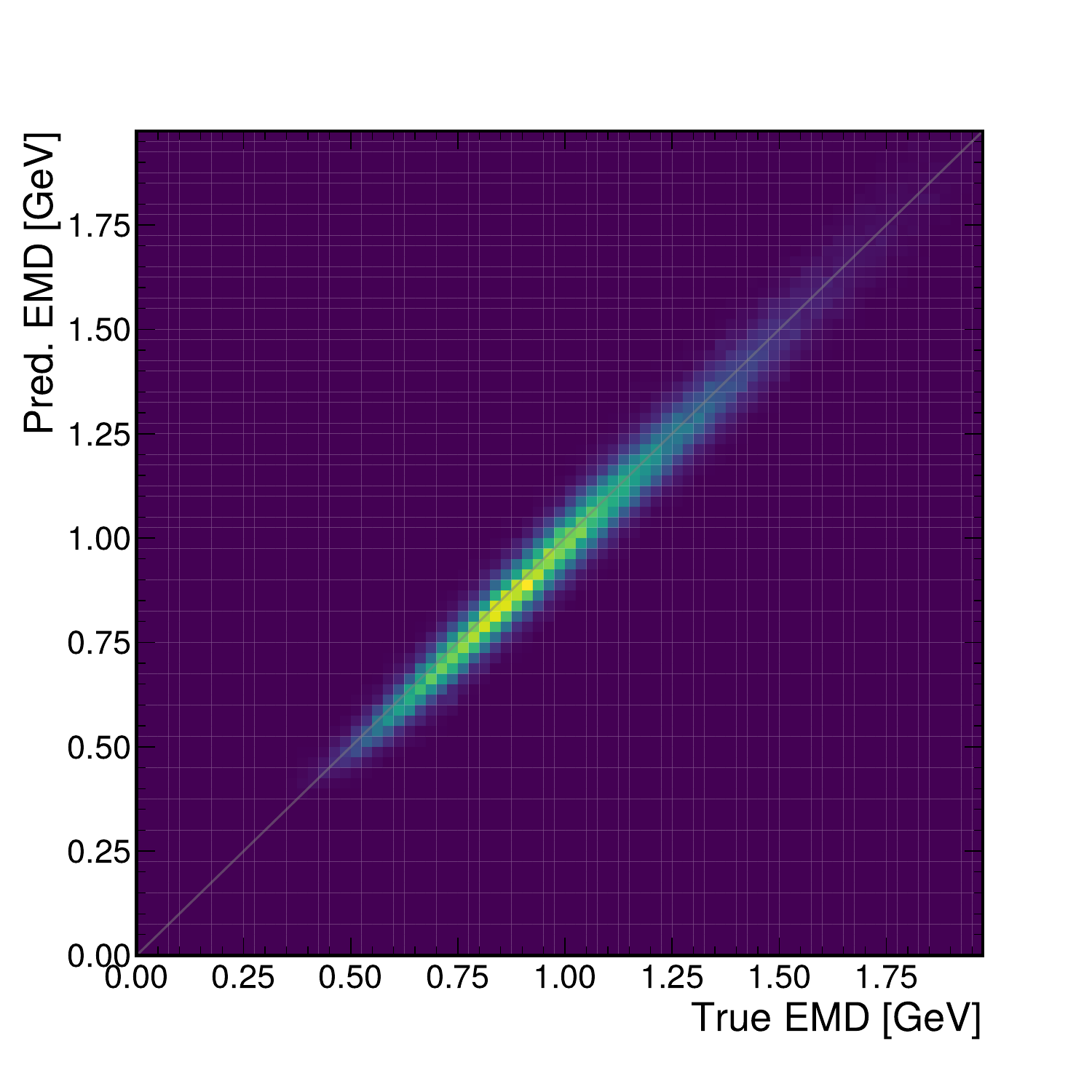}
    \caption{
        Performance of the EMD CNN with optimized hyperparameters.
        Distribution of the relative difference between the EMD CNN prediction and the true EMD (left)
        Predicted EMD as a function of true EMD (right). 
        The optimized hyperparameters correspond to 32 convolutional filters, kernel size of 5, 4 2D convolutional layers, 1 fully-connected layer with 256 neurons, and MSE loss function.}
    \label{fig:EMD_perf}
\end{figure}

Figure~\ref{fig:EMD_perf} shows the performance of this model in predicting the EMD.
On the left, the distribution of the relative difference between the predicted and true EMD validation dataset is shown.
There is a relatively small bias ($\mu_\text{EMD rel. diff.}=-1.3\%$) and the standard deviation $\sigma_\text{EMD rel. diff.} = 5.3\%$ shows good resolution.
With this optimized EMD CNN model, we can now define a differentiable EMD loss function for use in training the autoencoder.

\section{Differentiable EMD Loss}
\label{sec:aeEMD}

The ASIC encoder for the HGCAL is a CNN trained in \QKeras with quantization aware training.
The autoencoder is trained with the aforementioned trigger cell encoding (A).
The convolutional layer has 8 nodes which encode the input into a 16 by 9 bit latent space.
This layer has a kernel size of 3 and a stride of 2.
The decoder mirrors the encoder.
We quantify the performance of this autoencoder by computing the EMD after training and binning them in functions of the wafer trigger cell charges.
Optimal trainings of the autoencoder have EMD values approaching zero.

\begin{figure}[ht]
    \centering
    \includegraphics[width=0.45\columnwidth]{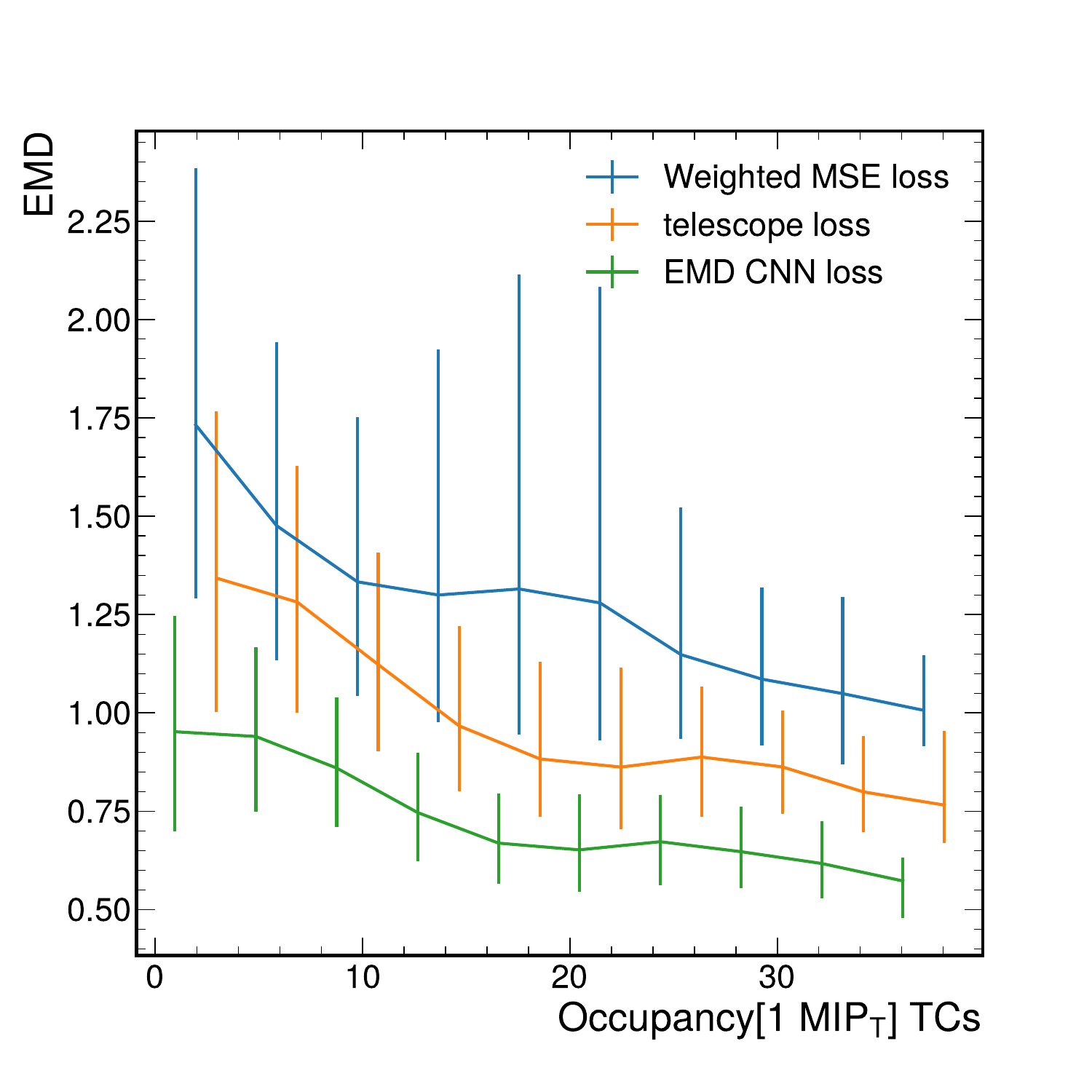}
    \includegraphics[width=0.45\columnwidth]{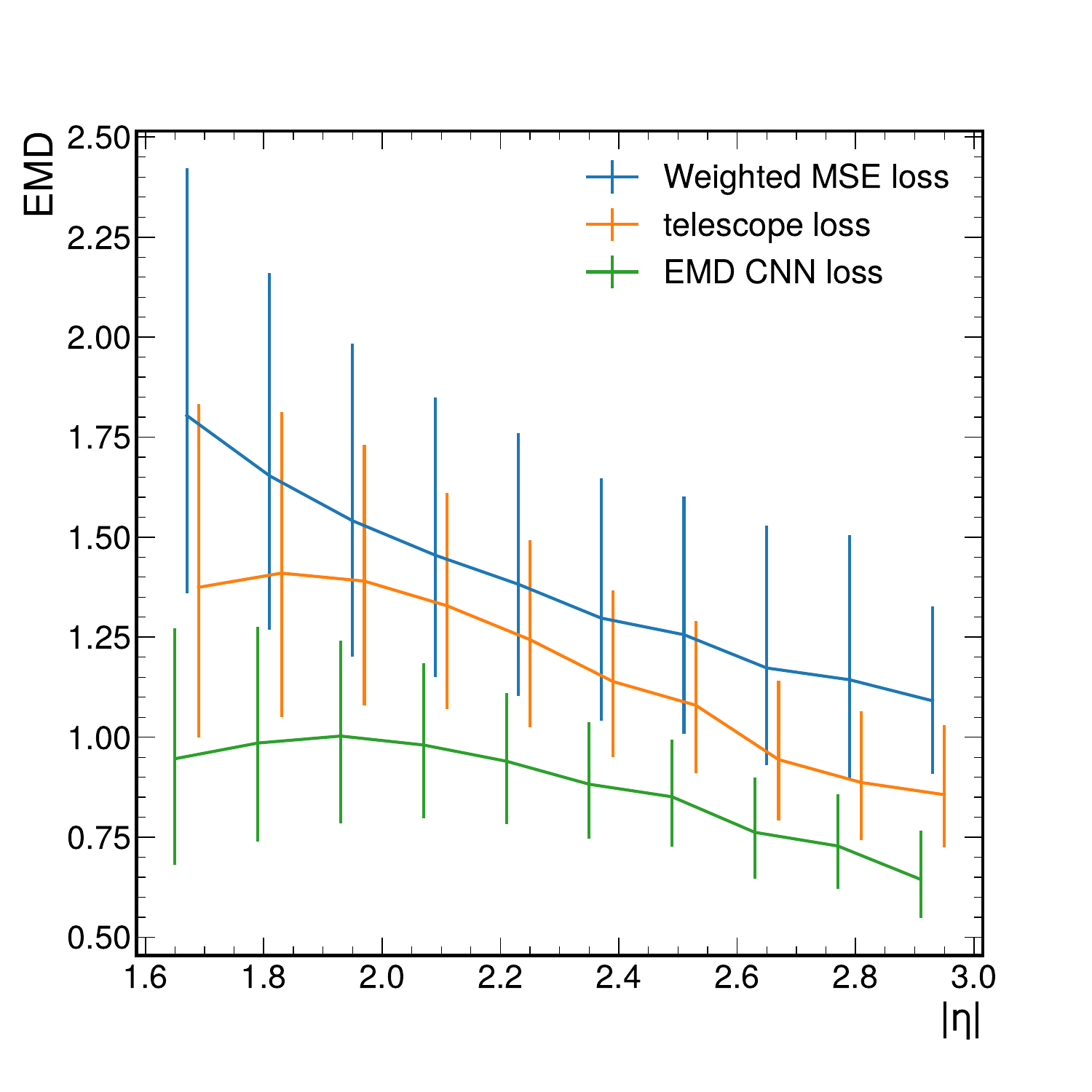}
    \caption{
        Median EMD for decoded HGCAL images versus the charge occupancy in the wafer (left) and wafer position (right).
        The vertical lines denote 68\% EMD intervals.
        Ideal autoencoder trainings have zero EMD between the input and output.
        Occupancy is defined as the number of TCs with signals exceeding one minimum ionizing particle divided by $\cosh\eta$.
    }
    \label{fig:ae}
\end{figure}

We now train this autoencoder with our EMD CNN loss.
We freeze the weights of an optimized EMD CNN, and use that as a loss function for back propagation. Comparing our results to weighted MSE and telescope MSE loss in Fig.~\ref{fig:ae} demonstrates an improved performance of the autoencoder with the EMD CNN loss. For instance for $1.8\le\eta\le 2.0$, the median EMD improves by 35\% with respect to the weighted MSE and by 28\% with respect to the telesecope MSE.

\section{Physics Performance}
\label{sec:physics}

\begin{figure}[ht]
    \centering
    \includegraphics[width=0.96\textwidth]{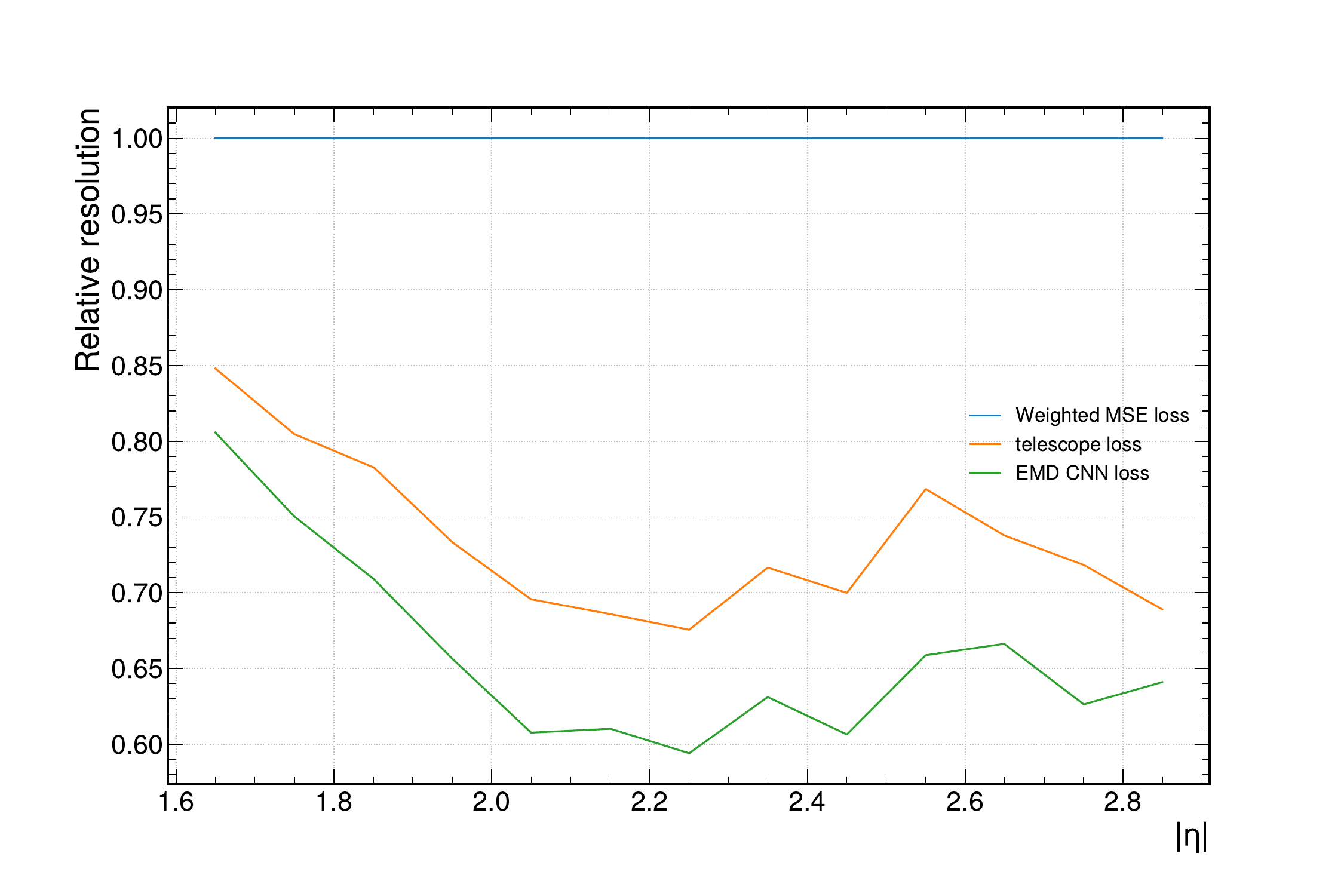}
    \caption{Performance relative to weighted MSE as a function of $\eta$.
        We define the performance as \RMSeff of $\pT^\text{L1T}/\pT^\text{gen}$, where \RMSeff is one standard deviation away from the mean.
        We divide the \RMSeff by the mean of distribution in that bin.}
    \label{fig:phys}
\end{figure}

While the task of the autoencoder is to reproduce the input wafers, our ultimate goal is to reconstruct electrons (and other particles) in the HGCAL in the L1T in real time.
With access to back-end FPGA resources, we can decode trigger cells, and cluster them to reconstruct trigger observables.
We evaluate the performance of the autoencoder by decoding and clustering single electron events for the three different tested loss functions.
To evaluate the resolution, we first bin the clusters by the pseudorapidity $\eta$ of the generated electron.
In this bin, we calibrate the energy of the cluster, by a multiplicative factor: to account for the L1T reading only half the layers, and an additive factor: to correct for pileup.
We then compute the effective RMS (\RMSeff) as the half-width of the shortest interval containing 68\% of the electron events.
We scale to the mean of the predicted \pT to obtain the resolution.
An ideal implementation of the ECON-T has zero \RMSeff.
The \RMSeff is shown as a function of $\eta$ in Fig.~\ref{fig:phys}.
By using the EMD CNN instead of the other loss functions, we find we achieve a better physics resolution for all values of $\eta$.
In particular, the effective resolution decreases by 13\% at $\eta=2.0$ when using the EMD CNN relative to the telescope MSE.

\section{Summary}
\label{sec:summary}

In this paper, we developed a convolutional neural network (CNN) to approximate the Earth mover's distance (EMD), a nondifferentiable and computationally intensive distance metric,  for two-dimensional images.
To demonstrate the utility of this approximation, we showed how it can be used to improve the performance of an on-detector data compression algorithm for the high-luminosity upgrade of the LHC (HL-LHC).

Specifically, we used the EMD CNN as a custom loss function for an on-detector encoder in the CMS experiment.
The optimized version of this neural network approximated the EMD with a resolution of 5\%.
This neural network is a fast-to-compute and natively differentiable approximation.
We used the EMD CNN as a loss function for the energy concentrator trigger ASIC autoencoder.
By training the autoencoder with the EMD CNN loss we improved the median EMD between input and output by 25\% with respect to the previous best optimization procedure introduced in prior work~\cite{ASIC}.
Finally, we evaluated the physics performance of the autoencoder in terms of the resolution of the transverse momentum of electrons by simulating the HL-LHC trigger path.
The EMD CNN loss improves the effective resolution by 13\% at a pseudorapidity $\eta=2.0$.

Beyond this particular application, the technique of approximating optimal transport functions using neural networks is applicable to many scientific tasks where it is important to minimize nondifferentiable or computationally intensive metrics. 
For example, this technique could be used to train an autoencoder to detect jet anomalies~\cite{steven}, cluster jets~\cite{Kitouni:2022qyr}, or mitigate the effects of pileup~\cite{gouskos2023optimal}.
In addition, a differentiable EMD approximation can be used to construct observables in searches for new physics~\cite{CrispimRomao:2020ejk} that can then be directly optimized for maximum sensitivity via backpropagation.
Finally, image- or point-cloud-based generative models often use EMD to evaluate the fidelity of generated data~\cite{Kansal:2021cqp,Kansal:2022spb,Mikuni:2022xry,Buhmann:2023bwk}, and using a differentiable approximation would enable direct minimization during training~\cite{deepemd2023}.
By directly optimizing EMD, better and more computationally efficient physics simulators can be trained, potentially enabling end-to-end detector optimization via differentiable programming~\cite{MODE:2022znx}.

\section*{Acknowledgments}
The authors would like to thank the CMS HGCAL Group for providing simulated module images for training, for input on the network optimization, and for feedback on this manuscript.
The authors would also like to thank the ECON-T ASIC team, particularly Cristian Gingu for slow control integration.

This research was supported by the Department of Energy (DOE), Office of Science, Office of Advanced Scientific Computing Research under the ``Real-time Data Reduction Codesign at the Extreme Edge for Science'' Project (DE-FOA-0002501).
JD was also supported by the DOE Office of Science, Office of High Energy Physics Early Career Research Program under award number DE-SC0021187 and the National Science Foundation (NSF) under award number 2117997 (A3D3).
RS was also supported by the UC San Diego Triton Research \& Experiential Learning Scholars (TRELS) program and an Undergraduate Summer Research Award.

Some neural network training was performed on the National Research Platform which is supported by NSF awards CNS-1730158, ACI-1540112, ACI-1541349, OAC-1826967, OAC-2112167, CNS-2100237, CNS-2120019, the University of California Office of the President, and the University of California San Diego's California Institute for Telecommunications and Information Technology/Qualcomm Institute.
Thanks to CENIC for the 100\,Gbps networks.

\section*{Author contributions statement}
RS conducted most of the training and evaluation.
JD and MP conceptualized some of the original ideas.
CH, DN, and CMS prepared the datasets and wrote the training framework.
CH, JH, DN, NT, and CMS all supervised the work.
All authors contributed to writing and reviewing the work and the manuscript.

\appendix
\section{Telescope MSE Details}
\label{sec:table}

Table~\ref{tab:2x2sums} shows the 36 2$\times$2 sums and corresponding weights used in the computation of the telesecope MSE loss described in Section~\ref{sec:losses}

\begin{table}[htpb]
    \centering
    \hfill
    \begin{tabular}{c|c}\footnotesize
        2$\times$2                       & Weight   \\
        \hline
        $\begin{bmatrix} q_{0} & q_{1} \\ q_{8} & q_{9}  \end{bmatrix}$ & 2.25/4   \\[10pt]
        $\begin{bmatrix} q_{3} & q_{4} \\ q_{11} & q_{12}  \end{bmatrix}$ & 1.5/4    \\[10pt]
        $\begin{bmatrix} q_{6} & q_{7} \\ q_{14} & q_{15}  \end{bmatrix}$ & 2.25/4   \\[10pt]
        $\begin{bmatrix} q_{10} & q_{11} \\ q_{18} & q_{19}  \end{bmatrix}$ & 1/4      \\[10pt]
        $\begin{bmatrix} q_{13} & q_{14} \\ q_{21} & q_{22}  \end{bmatrix}$ & 1/4      \\[10pt]
        $\begin{bmatrix} q_{17} & q_{18} \\ q_{25} & q_{26}  \end{bmatrix}$ & 1/4      \\[10pt]
        $\begin{bmatrix} q_{20} & q_{21} \\ q_{28} & q_{29}  \end{bmatrix}$ & 1.0833/4 \\[10pt]
        $\begin{bmatrix} q_{24} & q_{25} \\ q_{32} & q_{33}  \end{bmatrix}$ & 1.5/4    \\[10pt]
        $\begin{bmatrix} q_{28} & q_{29} \\ q_{35} & q_{39}  \end{bmatrix}$ & 1.1667/4 \\[10pt]
        $\begin{bmatrix} q_{32} & q_{33} \\ q_{36} & q_{37}  \end{bmatrix}$ & 1.5/4    \\[10pt]
        $\begin{bmatrix} q_{36} & q_{37} \\ q_{40} & q_{41}  \end{bmatrix}$ & 1.5/4    \\[10pt]
        $\begin{bmatrix} q_{40} & q_{41} \\ q_{44} & q_{45}  \end{bmatrix}$ & 2.25/4   \\[10pt]
    \end{tabular}
    \hfill
    \begin{tabular}{c|c}
        2$\times$2                       & Weight   \\
        \hline
        $\begin{bmatrix} q_{1} & q_{2} \\ q_{9} & q_{10}  \end{bmatrix}$ & 1.5/4    \\[10pt]
        $\begin{bmatrix} q_{4} & q_{5} \\ q_{12} & q_{13}  \end{bmatrix}$ & 1.5/4    \\[10pt]
        $\begin{bmatrix} q_{8} & q_{9} \\ q_{16} & q_{17}  \end{bmatrix}$ & 1.5/4    \\[10pt]
        $\begin{bmatrix} q_{11} & q_{12} \\ q_{19} & q_{20}  \end{bmatrix}$ & 1/4      \\[10pt]
        $\begin{bmatrix} q_{14} & q_{15} \\ q_{22} & q_{23}  \end{bmatrix}$ & 1.5/4    \\[10pt]
        $\begin{bmatrix} q_{18} & q_{19} \\ q_{26} & q_{27}  \end{bmatrix}$ & 1.0833/4 \\[10pt]
        $\begin{bmatrix} q_{21} & q_{22} \\ q_{29} & q_{30}  \end{bmatrix}$ & 1/4      \\[10pt]
        $\begin{bmatrix} q_{25} & q_{26} \\ q_{33} & q_{34}  \end{bmatrix}$ & 1/4      \\[10pt]
        $\begin{bmatrix} q_{29} & q_{30} \\ q_{39} & q_{43}  \end{bmatrix}$ & 1/4      \\[10pt]
        $\begin{bmatrix} q_{33} & q_{34} \\ q_{37} & q_{38}  \end{bmatrix}$ & 1/4      \\[10pt]
        $\begin{bmatrix} q_{37} & q_{38} \\ q_{41} & q_{42}  \end{bmatrix}$ & 1/4      \\[10pt]
        $\begin{bmatrix} q_{41} & q_{42} \\ q_{45} & q_{46}  \end{bmatrix}$ & 1.5/4    \\[10pt]
    \end{tabular}
    \hfill
    \begin{tabular}{c|c}
        2$\times$2                       & Weight   \\
        \hline
        $\begin{bmatrix} q_{2} & q_{3} \\ q_{10} & q_{11}  \end{bmatrix}$ & 1.5/4    \\[10pt]
        $\begin{bmatrix} q_{5} & q_{6} \\ q_{13} & q_{14}  \end{bmatrix}$ & 1.5/4    \\[10pt]
        $\begin{bmatrix} q_{9} & q_{10} \\ q_{17} & q_{18}  \end{bmatrix}$ & 1/4      \\[10pt]
        $\begin{bmatrix} q_{12} & q_{13} \\ q_{20} & q_{21}  \end{bmatrix}$ & 1/4      \\[10pt]
        $\begin{bmatrix} q_{16} & q_{17} \\ q_{24} & q_{25}  \end{bmatrix}$ & 1.5/4    \\[10pt]
        $\begin{bmatrix} q_{19} & q_{20} \\ q_{27} & q_{28}  \end{bmatrix}$ & 1.1667/4 \\[10pt]
        $\begin{bmatrix} q_{22} & q_{23} \\ q_{30} & q_{31}  \end{bmatrix}$ & 1.5/4    \\[10pt]
        $\begin{bmatrix} q_{26} & q_{27} \\ q_{34} & q_{35}  \end{bmatrix}$ & 1.1667/4 \\[10pt]
        $\begin{bmatrix} q_{30} & q_{31} \\ q_{43} & q_{47}  \end{bmatrix}$ & 1.5/4    \\[10pt]
        $\begin{bmatrix} q_{34} & q_{35} \\ q_{38} & q_{39}  \end{bmatrix}$ & 1.1667/4 \\[10pt]
        $\begin{bmatrix} q_{38} & q_{39} \\ q_{42} & q_{43}  \end{bmatrix}$ & 1./4     \\[10pt]
        $\begin{bmatrix} q_{42} & q_{43} \\ q_{46} & q_{47}  \end{bmatrix}$ & 1.5/4
    \end{tabular}
    \hfill
    \caption{Arrays of 2$\times$2 trigger cells summed for $L_{2{\times}2}$, and corresponding weight factors to account for the number of times constituent cells show up in 2$\times$2 sums.}
    \label{tab:2x2sums}
\end{table}

\section{EMD CNN Hyperparameter Optimization}
\label{sec:loss}

The eight best configurations searched when optimizing the hyperparameters are presented in Table~\ref{tab:hyper}.
Despite performing 15\% worse than Configuration 1 on predicting EMD, Configuration 4 performs the best when used as a loss function for the autoencoder. We thus use Configuration 4 throughout the paper. 

\begin{table}[htpb]
    \begin{tabular}{l|cccccccc}
        Hyperparameter configuration   & 1              & 2     & 3     & 4     & 5     & 6     & 7     & 8     \\\hline
        Number of filters              & 64             & 32    & 128   & 32    & 128   & 32    & 128   & 128   \\
        Kernel size                    & 5              & 5     & 5     & 5     & 5     & 5     & 5     & 3     \\
        Number of 2D conv. layers      & 4              & 4     & 3     & 3     & 4     & 3     & 4     & 4     \\
        Number of dense layers         & 1              & 1     & 1     & 1     & 1     & 1     & 1     & 1     \\
        Number of dense layer neurons  & 32             & 32    & 64    & 256   & 256   & 128   & 32    & 256   \\\hline
        $\sigma_\text{EMD rel. diff.}$ & \textbf{0.067} & 0.069 & 0.071 & \textbf{0.078} & 0.078 & 0.079 & 0.083 & 0.086 \\
        Mean performance on AE         & 3.27           & 3.39  & 3.41  & \textbf{1.03}  & 3.43 & 2.01  & 3.23 & 1.10\\
    \end{tabular}
    \label{tab:hyper}
    \caption{The eight best hyperparameter configurations found when optimizing the EMD CNN.}
\end{table}

\clearpage
\section*{References}
\bibliography{bibliography}
\bibliographystyle{cms_unsrt}
\end{document}